\documentclass{llncs}

\usepackage{listings}

\usepackage{color}
\usepackage{url}
\usepackage[colorlinks,linkcolor=black,urlcolor=black,citecolor=black]{hyperref}

\usepackage{amsmath}
\usepackage{varioref}
\usepackage{amssymb}
\usepackage[algoruled,linesnumbered]{algorithm2e}
\usepackage{textcomp}
\usepackage{graphicx}
\usepackage{palatino}
\usepackage{mathrsfs}
\usepackage{dsfont}
\usepackage{multicol}

\usepackage{caption}
\usepackage{subcaption}

\usepackage{tikz}
\usetikzlibrary{arrows,automata}
\usetikzlibrary{decorations.pathmorphing}
\tikzset{initial text={}}

\SetKwBlock{SubAlgoBlock}{}{end}

\SetKw{Var}{var}
\SetKw{Payload}{payload}
\SetKw{Query}{query}
\SetKw{Update}{update}
\SetKw{Receive}{on receive}
\SetKw{Broadcast}{broadcast}

\SetKw{True}{true}
\SetKw{False}{false}

\begin{document}

\title{Brief Announcement: Update Consistency in Partitionable Systems}
\author{Matthieu Perrin \and Achour Most\'efaoui \and Claude Jard}
\institute{LINA -- University of Nantes, 2 rue de la Houssini\`ere, 44322 Nantes
  Cedex 3, France \\
  \email{matthieu.perrin@univ-nantes.fr}\\
  \email{achour.mostefaoui@univ-nantes.fr}\\
  \email{claude.jard@univ-nantes.fr}
}

\sloppy
\maketitle
Data replication is essential to ensure reliability, availability and 
fault-tolerance of massive distributed applications
over large scale systems such as the Internet. However, these systems are prone to partitioning,
which by Brewer's CAP theorem \cite{gilbert2002brewer} 
makes it impossible to use a strong consistency criterion like
atomicity. Eventual consistency \cite{vogels2008eventually}
guaranties that all replicas eventually converge to a common
state when the participants stop updating. However, it fails to fully
specify shared objects and requires additional non-intuitive and error-prone
distributed specification techniques, that must take into account all possible 
concurrent histories of updates to specify this common state
\cite {burckhardt2014replicated}.
This approach, that can lead to specifications as complicated as the 
implementations themselves, is limited by a more serious issue.
The concurrent specification of objects uses the notion of \emph{concurrent events}.
In message-passing 
systems, two events are concurrent if they are enforced by different processes and
each process
enforced its event before it received the notification message from the other
process. In other words,
the notion of concurrency depends on the implementation of the object, not on its
specification.
Consequently, the final user may not know if two events are concurrent without
explicitly tracking
the messages exchanged by the processes. A specification should be independent of
the system on
which it is implemented.

We believe that an object should be totally specified by two facets: 
its abstract data type, that characterizes its sequential executions, 
and a consistency criterion, that defines how it is supposed to behave 
in a distributed environment.
Not only sequential specification helps repeal the problem of intention, it also
allows to use the well studied and understood notions of languages and automata.
This makes possible to apply all the tools developed for sequential systems, from
their simple definition using structures and classes to the
most advanced techniques like model checking and formal verification. 

\begin{figure}[t]
  \centering
  \begin{subfigure}[b]{0.30\textwidth}
    \centering
    \begin{tikzpicture}[scale=0.80, transform shape]
      \draw     (0.5,1) -- (4.5,1) ;
      \draw     (1,1)    node{$\bullet$} ;
      \draw     (1,1)    node[above]{$\text{I}(1)$} ;
      \draw     (2,1)  node{$\bullet$} ;
      \draw     (2,1)  node[above]{$\text{D}(2)$} ;
      \draw     (3,1)  node{$\bullet$} ;
      \draw     (3,1)  node[above]{$\text{R}_{\{1\}}$} ;
      \draw     (4,1)  node{$\bullet$} ;
      \draw     (4,1)  node[above]{$\text{R}_{\{2\}}^\omega$} ;
      
      \draw     (0.5,0.5) -- (4.5,0.5) ;
      \draw     (1,0.5)    node{$\bullet$} ;
      \draw     (1,0.5)    node[below]{$\text{I}(2)$} ;
      \draw     (2,0.5)  node{$\bullet$} ;
      \draw     (2,0.5)  node[below]{$\text{D}(1)$} ;
      \draw     (3,0.5)    node{$\bullet$} ;
      \draw     (3,0.5)    node[below]{$\text{R}_{\emptyset}$} ;
      \draw     (4,0.5)    node{$\bullet$} ;
      \draw     (4,0.5)    node[below]{$\text{R}_{\{1\}}^\omega$} ;
    \end{tikzpicture}
      \caption{Not EC and not UC}
      \label{figure:rien}
  \end{subfigure}
  ~~
  \begin{subfigure}[b]{0.30\textwidth}
    \centering
    \begin{tikzpicture}[scale=0.80, transform shape]
      \draw     (0.5,1) -- (4.5,1) ;
      \draw     (1,1)    node{$\bullet$} ;
      \draw     (1,1)    node[above]{$\text{I}(1)$} ;
      \draw     (2,1)  node{$\bullet$} ;
      \draw     (2,1)  node[above]{$\text{D}(2)$} ;
      \draw     (3,1)  node{$\bullet$} ;
      \draw     (3,1)  node[above]{$\text{R}_{\{1\}}$} ;
      \draw     (4,1)  node{$\bullet$} ;
      \draw     (4,1)  node[above]{$\text{R}_{\{1, 2\}}^\omega$} ;
      
      \draw     (0.5,0.5) -- (4.5,0.5) ;
      \draw     (1,0.5)    node{$\bullet$} ;
      \draw     (1,0.5)    node[below]{$\text{I}(2)$} ;
      \draw     (2,0.5)  node{$\bullet$} ;
      \draw     (2,0.5)  node[below]{$\text{D}(1)$} ;
      \draw     (3,0.5)    node{$\bullet$} ;
      \draw     (3,0.5)    node[below]{$\text{R}_{\emptyset}$} ;
      \draw     (4,0.5)    node{$\bullet$} ;
      \draw     (4,0.5)    node[below]{$\text{R}_{\{1, 2\}}^\omega$} ;
    \end{tikzpicture}
      \caption{EC but not UC}
      \label{figure:ec}
  \end{subfigure}
  ~~
  \begin{subfigure}[b]{0.30\textwidth}
    \centering
    \begin{tikzpicture}[scale=0.80, transform shape]
      \draw     (0.5,1) -- (4.5,1) ;
      \draw     (1,1)    node{$\bullet$} ;
      \draw     (1,1)    node[above]{$\text{I}(1)$} ;
      \draw     (2,1)  node{$\bullet$} ;
      \draw     (2,1)  node[above]{$\text{D}(2)$} ;
      \draw     (3,1)  node{$\bullet$} ;
      \draw     (3,1)  node[above]{$\text{R}_{\{1\}}$} ;
      \draw     (4,1)  node{$\bullet$} ;
      \draw     (4,1)  node[above]{$\text{R}_{\{1\}}^\omega$} ;
      
      \draw     (0.5,0.5) -- (4.5,0.5) ;
      \draw     (1,0.5)    node{$\bullet$} ;
      \draw     (1,0.5)    node[below]{$\text{I}(2)$} ;
      \draw     (2,0.5)  node{$\bullet$} ;
      \draw     (2,0.5)  node[below]{$\text{D}(1)$} ;
      \draw     (3,0.5)    node{$\bullet$} ;
      \draw     (3,0.5)    node[below]{$\text{R}_{\emptyset}$} ;
      \draw     (4,0.5)    node{$\bullet$} ;
      \draw     (4,0.5)    node[below]{$\text{R}_{\{1\}}^\omega$} ;
    \end{tikzpicture}
      \caption{EC and UC}
      \label{figure:uc}
  \end{subfigure}
  \caption{Three histories for a set of integers, with different
    consistency criteria. An event
    labeled $\omega$ is repeated infinitely often.}
  \label{figure:histories}
\end{figure}
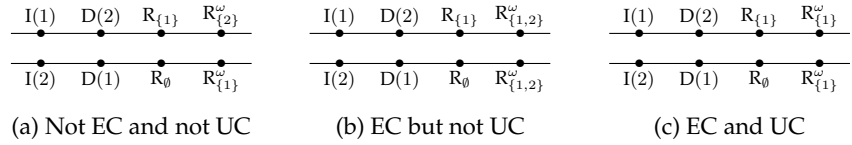

%\paragraph{Update Consistency.} 
Eventual consistency (EC) imposes no constraint on the 
convergent state, that very few depends on the sequential specification.
For example, an implementation that ignores all the updates is eventually consistent, 
as all replicas converge to the initial state. 
We propose a new consistency criterion, update consistency (UC), in which 
the convergent state must be obtained by a total ordering
of the updates, that contains the sequential order of each process.
Another equivalent way to approach it is that,
if the number of updates is finite, it is possible to remove a finite number of
queries such that the remaining history is sequentially consistent.
Unlike Fig. \ref{figure:rien}, Fig. \ref{figure:ec} presents an eventually 
consistent history, as both processes read $\{1, 2\}$ once they have converged.
However, it is not update consistent: in any linearization of the updates, a deletion
must appear as the last update, so this history cannot converge to state $\{1, 2\}$. 
State $\{1\}$ is possible because the updates can be done 
in the order $\text{I}(2), \text{D}(1), \text{I}(1), \text{D}(2)$, so
Fig. \ref{figure:uc}, is update consistent.
As update consistency is strictly stronger than eventual consistency, 
an update consistent object can always be used instead of its eventually consistent counterpart.

%\paragraph{Generic construction.}
We can prove that update consistency is universal, in the sense that every object
has an update consistent implementation in a partitionable system, where any number of crashes are allowed. 
The principle is to build a total order on the updates
on which all the participants agree, and then to rewrite the history \emph{a posteriori}
so that every replica of the object eventually reaches the state corresponding to the
common sequential history. Any strategy to build the total order on the updates
would work. For example, this order can be built from a timestamp made of a Lamport's
clock \cite{lamport1978time} and the id of the process that performed it. 
The genericity of the proposed algorithm is very important because it 
may give a
substitute to composability. 
Composability is an important property of consistency criteria because 
it allows to program
in a modular way, but it is very difficult to achieve for consistency 
criteria. A same algorithm that pilots several objects during a same 
execution allows this execution to be update consistent.
This universality result allows to imagine automatic compilation 
techniques that compose
specifications instead of implementations.

\bibliographystyle{splncs}
\bibliography{brief}

\end{document}